\documentclass[%
 reprint,
 amsmath,amssymb,
 aps,
]{revtex4-2}

\usepackage{graphicx}
 \usepackage[all,cmtip]{xy}
\usepackage{dcolumn}
\usepackage{bm}

\newcommand{\overbar}[1]{\mkern 1.5mu\overline{\mkern-1.5mu#1\mkern-1.5mu}\mkern 1.5mu}

\def\TT{{T\overbar{T}}}

\newcommand{\TTbar}{T\overbar{T}}
\newcommand{\Det}[1]{\text{Det}\left[{#1}\right]}

\newcommand{\de}{\text{d}}
\newcommand{\psib}{\overbar{\psi}}

\begin{document}


\title{\boldmath Root-${\TT}$ Deformations in Two-Dimensional Quantum Field Theories}

\author{Christian Ferko}\email{caferko@ucdavis.edu}
\affiliation{%
Center for Quantum Mathematics and Physics (QMAP),
Department of Physics \& Astronomy, University of California, Davis, CA 95616, USA.}

\author{Alessandro Sfondrini}\email{alessandro.sfondrini@unipd.it, IBM-Einstein Fellow}
 \affiliation{Dipartimento di Fisica e Astronomia, Universit\`a degli Studi di Padova, \& Istituto Nazionale di Fisica Nucleare, Sezione di Padova,
via Marzolo 8, 35131 Padova, Italy.\\}
\affiliation{Institute for Advanced Study\\
Einstein Drive, Princeton, New Jersey, 08540 USA}

\author{Liam Smith}\email{liam.smith1@uq.net.au}
 \author{Gabriele Tartaglino-Mazzucchelli}%
 \email{g.tartaglino-mazzucchelli@uq.edu.au}
\affiliation{%
School of Mathematics and Physics, University of Queensland,
St Lucia, Brisbane, Queensland 4072, Australia.}%

\date{\today}

\begin{abstract}
In this letter we introduce a one-parameter deformation of two-dimensional quantum field theories generated by a non-analytic operator which we call Root-$\TTbar$.
For a conformal field theory, the operator coincides with the square-root of the $\TTbar$ operator.
More generally, the operator is defined so that classically it is marginal and generates a flow which commutes with the $\TTbar$-flow.
Intriguingly, the Root-$\TTbar$ flow is closely related to the ModMax theory recently constructed by Bandos, Lechner, Sorokin and Townsend.

\end{abstract}

\maketitle


\section{Introduction and Summary}
The so-called \textit{$\TTbar$ operator} is a universal composite operator defined  from the stress-energy tensor~$T^{\mu\nu}$ of any two-dimensional quantum field theory (QFT) as~\cite{Zamolodchikov:2004ce}
\begin{subequations}
\label{eq:TTbardef}
\begin{eqnarray}
 {\cal{O}}&=&\int\de^2z\, O(z,\bar{z})
     \,,\\
     O(z,\bar{z})
     &=&
     - \Det{T^{\mu\nu}}
     =
     \frac{1}{2}\left(T^{\mu\nu}T_{\mu\nu}-T_\mu{}^\mu T_\nu{}^\nu\right)\,.~~~~~~
\end{eqnarray}
\end{subequations}
In a conformal field theory (CFT) the trace of the stress-energy tensor vanishes and we have, in light-cone coordinates, $O(z,\bar{z})=T(z)\overline{T}(\bar{z})$. This is however not the case for a generic QFT, making ``$\TTbar$'' somewhat of a misnomer.
Given a two-dimensional QFT defined by a ``seed'' action~${\cal S}_0$, the operator ${\cal O}$ can be used to define a one-parameter family of theories~${\cal S}_{\lambda}$~\cite{Smirnov:2016lqw,Cavaglia:2016oda}
\begin{equation}
    \partial_\lambda {\cal S}_\lambda = {\cal{O}}_\lambda\,,\qquad {\cal S}_{\lambda=0}={\cal S}_0\,.
\end{equation}
Because $O$ has scaling dimension four, the resulting flow is \textit{irrelevant} in the sense of the renormalization group (RG). This is counter-intuitive in the usual RG picture, where one typically considers \textit{relevant} or \textit{marginal} (\textit{i.e.}~conformal) perturbations. Nonetheless, the $\TTbar$ flow has many remarkable properties. Applied to a CFT, it breaks scale-invariance but preserves infinitely-many commuting charges, and it results in a family of integrable QFTs (IQFTs). If ${\cal S}_0$ is not conformal, but integrable, the flow still preserves integrability. In fact, the flow is so well-behaved that it is possible to express the finite-volume spectrum of the deformed theory~${\cal S}_\lambda$ in closed form starting from the spectrum of the seed theory~${\cal S}_0$.
These properties allowed for a detailed study of $\TTbar$-deformed theories which has revealed and is revealing many surprising connections to string theory~\cite{Cavaglia:2016oda,Baggio:2018gct,Frolov:2019nrr}, holography~\cite{McGough:2016lol,Giveon:2017nie,Dubovsky:2017cnj}, two-dimensional gravity~\cite{Dubovsky:2018bmo}, geometry~\cite{Conti:2018tca}, random geometry~\cite{Cardy:2018sdv}, and more
--- see \cite{Jiang:2019epa,Ferko:2021loo} for reviews and references on $\TTbar$.

\begin{figure}
\centering
\includegraphics{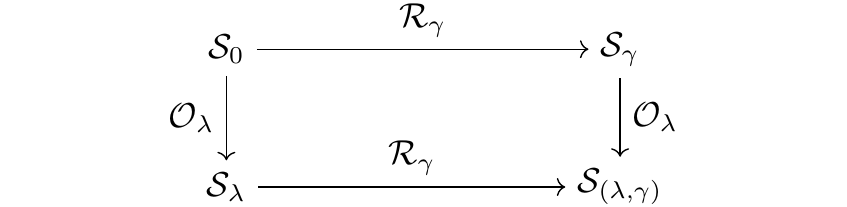}
\caption{We require that first deforming ${\cal S}_0$ by $\TTbar$ with ${\cal{O}}_\lambda$ and then deforming the result by Root-$\TTbar$ with ${\cal{R}}_\gamma$ is the same as deforming ${\cal S}_0$ by ${\cal{R}}_\gamma$ and then by ${\cal{O}}_\lambda$.}
\label{fig:commute}
\end{figure}

In this letter we propose a new  ``square-root'', $\TTbar$-like deformation. By this we mean a \textit{marginal} deformation defined by $\partial_\gamma {\cal S}_\gamma={\cal{R}}_\gamma$ which, when applied to a two-dimensional conformal field theory (CFT$_2$) ${\cal S}_0$, amounts to deforming by a non-analytic operator 
\begin{subequations}
\label{Root-TTbar_CFT}
\begin{eqnarray}
{\cal{R}}&=&\int\de^2z\, R(z,\bar{z})\,,
\\
R(z,\bar{z})
&=&
\sqrt{T(z)\overline{T}(\bar{z})}\,,
\qquad\text{(for a CFT).}
\end{eqnarray}
\end{subequations}
Despite the square-root, this promises to be a well-behaved expression in a CFT$_2$ because the operator-product expansion between a chiral and an anti-chiral field (non-analytic as they may be) is regular.
This deformation was recently considered in \cite{Rodriguez:2021tcz,Bagchi:2022nvj} to study  the BMS$_3$ symmetry  of  ultra/non-relativistic limits of a CFT$_2$.
It is however not obvious how such an expression should be generalised when ${\cal S}_0$ is \textit{not} conformal. We do this by requiring that the Root-$\TTbar$ flow commutes with the ordinary $\TTbar$ flow, \textit{i.e.}\ that we can construct a \textit{two-parameter} action ${\cal S}_{(\lambda,\gamma)}$ so that the diagram in Figure~\ref{fig:commute} commutes. As we shall explain in the conclusions, this choice will make it easier to study a vaster class of theories.
We will show that, classically, this can be obtained by setting
\begin{equation}
\label{eq:rootttbar}
    R(z,\bar{z})=\sqrt{\frac{1}{2}T^{\mu\nu}T_{\mu\nu}-\frac{1}{4}T_{\mu}{}^\mu T_{\nu}{}^\nu}=
    \sqrt{-\text{Det}\big[t^{\mu\nu}\big]}\,,
\end{equation}
where
$
    t^{\mu\nu}=T^{\mu\nu}-\frac{1}{2}g^{\mu\nu}T_\rho{}^\rho
$
is the traceless part of the stress-energy tensor $T^{\mu\nu}$.
We will construct the explicit two-parameter action ${\cal S}_{(\lambda,\gamma)}$ for a theory of $N$ scalar fields. This turns out to be related to the action for the unique conformal- and Electro-Magnetic duality-invariant deformation of Maxwell theory, the \textit{ModMax} theory~\cite{Bandos:2020jsw,Bandos:2020hgy}
(see also \cite{Kosyakov:2020wxv,Sorokin:2021tge}). ModMax is a four-dimensional model, and it depends on a single marginal parameter~$\gamma$
\begin{equation}
\label{eq:MM}
    \mathcal{S}^{\text{MM}}_{\gamma}=\int \de^4x\left[-\frac{\text{ch}\gamma}{4}F^2+\frac{\text{sh}\gamma}{4}\sqrt{(F^2){}^2+(F\widetilde{F}){}^2}\right],
\end{equation}
where $F$ indicates the electromagnetic field strength and $\tilde{F}$ its dual.
Eq.~\eqref{eq:MM} admits a $\TTbar$-like deformation in terms of a Born-Infeld action with parameter~$\lambda$. 
The resulting two parameter, four-dimensional action ${\cal S}_{(\lambda,\gamma)}$ is the solution of two commuting flow-equations \cite{Conti:2018jho,Babaei-Aghbolagh:2022uij,Ferko:2022iru}. The flows we introduce here precisely generate, and extend, the reduction to two dimensions of the ModMax-Born-Infeld action recently obtained in Ref.~\cite{Conti:2022egv}.
 After discussing the generalisation of our construction to actions involving Fermions, we will comment on the implications of this relation and on several immediate directions of study.

\paragraph*{Note added:}
On June $22^{\text{nd}}$ 2022, the day this work appeared for the first time on the arXiv, the Authors of~\cite{Conti:2022egv} released a second version of their manuscript where they independently showed that the operator \eqref{eq:rootttbar} generates the $\gamma$-flow of the Lagrangian \eqref{N_scalars_modBI} in the particular case of $N$ Bosons with seed Lagrangian \eqref{eq:seedL} when $G_{ij}(\phi)=\delta_{ij}$ and $B_{ij}(\phi)=0$.
On June $25^{\text{th}}$, Ref.~\cite{Babaei-Aghbolagh:2022kfz} appeared and presented an exploration of the relation of Root-$\TTbar$ and ModMax theories.

\section{Root-$\TTbar$ Flow for $N$ Scalars}
We start by considering a rather general Lagrangian density for $N$ real scalars $\phi^i$, of the form
\begin{equation}
\label{eq:seedL}
    L=\frac{1}{2}\Big(
    G_{ij}(\phi)g^{\mu\nu}
    +B_{ij}(\phi)\varepsilon^{\mu\nu}\Big)\partial_\mu\phi^i\partial_\nu\phi^j-V(\phi)\,,
\end{equation}
where $g^{\mu\nu}$ is the two-dimensional inverse metric (which we take to be Minkowski),
$\varepsilon^{\mu\nu}$ is the Levi-Civita tensor, and
$i=1,\cdots,N$. 
This ansatz encompasses many QFTs of interest including vector models, non-linear sigma models 
(where $G_{ij}$ and $B_{ij}$ represent the target-space metric and Kalb-Ramond fields, respectively),
and Wess-Zumino-Witten models, but it is not closed under Root-$\TTbar$ deformations. In fact, terms with a higher number of derivatives will appear along the flow, as it can be seen recalling that the (Hilbert) stress-energy tensor~is
\begin{equation}
T_{\mu\nu}=-2\frac{\partial L}{\partial g^{\mu\nu}}+g_{\mu\nu}L\,.
\end{equation}
To  write the general ansatz for a flow driven by the operator $R$~\eqref{eq:rootttbar}, we list the tensor structures that may arise from the stress-energy tensor associated to~\eqref{eq:seedL} and from its powers. As $g^{\mu\nu}$ only appears in the first term of~\eqref{eq:seedL}, we consider
\begin{equation}
    (X_1)_{\mu}{}^\nu=G_{ij}\partial_\mu\phi^i\partial^\nu\phi^j,\quad
    (X_2)_{\mu}{}^\nu=(X_1)_{\mu}{}^\rho(X_1)_{\rho}{}^\nu
    .
\end{equation}
In principle, we could define $(X_3)_{\mu}{}^\nu$, $(X_4)_{\mu}{}^\nu$, \textit{etc.}, in an obvious way. However, in the Lagrangian only the trace of these tensors may appear, and for any $2\times2$ matrix the only independent trace-invariants are $x_1=(X_1)_{\mu}{}^\mu$ and $x_2=(X_2)_{\mu}{}^\mu$, due to identities such as
\begin{equation}
\label{eq:traceids}
2x_3=3x_1x_2-x_1^3,\quad
2x_4=2x_1^2x_2-x_1^4+x_2^2\,,
\end{equation}
with $x_n=(X_n)_\mu{}^\mu$.
Hence the most general Lagrangian will depend on $x_1$, $x_2$, and possibly on a term $x_0$ which does not couple to $g^{\mu\nu}$, $L_{\gamma}=L_{\gamma}(x_0,x_1,x_2)$. For such a Lagrangian
\begin{equation}
\label{eq:Tmunuderivatives}
    T_{\mu\nu}=-2 (X_1)_{\mu\nu}\frac{\partial L_\gamma}{\partial x_1}
    -4(X_2)_{\mu\nu}\frac{\partial L_{\gamma}}{\partial x_2}+g_{\mu\nu}L_{\gamma}\,.
\end{equation}
Computing the contractions $T_{\mu\nu}T^{\mu\nu}$ and $(T_\mu{}^\mu)^2$, and simplifying the resulting expression with the help of the trace identities~\eqref{eq:traceids}, we can evaluate the square of the operator~$R$ of Eq.~\eqref{eq:rootttbar} as
\begin{equation}
\label{eq:TTminusT2}
    R^2
    =\left(2x_2-x_1^2\right)\left(\frac{\partial L_{\gamma}}{\partial x_1}+2x_1\frac{\partial L_{\gamma}}{\partial x_2}\right)^2.
\end{equation}
This expression is surprisingly nice in more than one way. Firstly, the partial derivatives appear in a perfect square so that, away from $\partial_{x_1}L_{\gamma}=-2 x_1 \partial_{x_{2}}L_{\gamma}$, the flow equation for the Root-$\TTbar$ operator~\eqref{eq:rootttbar} will be \textit{linear},
\begin{equation}
\label{eq:rotttflow}
    \frac{\partial}{\partial\gamma} L_\gamma= R_\gamma=\pm\sqrt{2x_{2}-x_{1}^2 }\left(\frac{\partial L_\gamma}{\partial x_1}+2x_1\frac{\partial L_\gamma}{\partial x_2}\right)\,.
\end{equation}
Both choices of the branch can be recovered by flipping the sign of~$\gamma$. As $2x_{2}=x_{1}^2$ gives a fixed point of~\eqref{eq:rotttflow}, the branch cannot change along the flow.
Additionally, the flow equation will not depend explicitly on~$L$, so that the dependence on~$x_0$ will affect the flow \textit{only through its initial conditions}. Finally, \eqref{eq:rotttflow} can be integrated,
\begin{equation}
\label{eq:Lgamma}
    L_\gamma\left(x_0,x_1,x_2\right)= L_{\gamma=0}\left(x_0,x_1^{(\gamma)},x_2^{(\gamma)}\right)\,,
\end{equation}
where
\begin{subequations}
\label{eq:X1X2}
\begin{eqnarray}
x_1^{(\gamma)} &=& \text{ch} ( \gamma )\, x_1 \pm \text{sh} ( \gamma ) \sqrt{2 x_2 - x_1^2} \, , 
\\
x_2^{(\gamma)} &=&  \text{ch} ( 2 \gamma )\,x_2 \pm \text{sh} ( 2 \gamma )\, x_1 \sqrt{2 x_2 - x_1^2} \, ,
\end{eqnarray}
\end{subequations}
and $x_0$ is not deformed. Note that the combination
\begin{equation}
\label{eq:constant}
    \big(x_1^{(\gamma)}\big)^2-x_2^{(\gamma)}=x_1^2-x_2\,,
\end{equation}
is constant along the flow. In fact, an action which depends on only this combination and through~$x_0$ is a fixed-point of this flow. We also note that~\eqref{eq:X1X2} has the property
\begin{align}
\label{eq:trace_unchanged}
    x_1 \frac{\partial L_{\gamma}}{\partial x_1} + 2 x_2 \frac{\partial L_{\gamma}}{\partial x_2} = x_1^{(\gamma)} \frac{\partial L_{\gamma}}{\partial x_1^{(\gamma)}} + 2 x_2^{(\gamma)} \frac{\partial L_{\gamma}}{\partial x_2^{(\gamma)}} \, ,
\end{align}
and that the trace of the stress-energy tensor for $L_{\gamma}$~\eqref{eq:Tmunuderivatives} vanishes if and only if
\begin{equation}
    \frac{1}{2}T_{\mu}{}^{\mu}
    =
    L_\gamma - \left( x_1\frac{\partial L_\gamma}{\partial x_1}+2
    x_2\frac{\partial L_\gamma}{\partial x_2} \right) = 0\,.
\end{equation}
Eq.~\eqref{eq:trace_unchanged} therefore implies that, if the stress-energy tensor 
for a seed theory $L_0$ is traceless, then the same is true for the deformed theory $L_\gamma$. Importantly, this confirms that the operator $R_\gamma$ defines a classically marginal deformation.

Note that, in the special case of $N$ free Bosons, it holds
\begin{equation}\label{eq:Lgamma_bosons}
    L_\gamma=
    \frac{\text{ch}\gamma}{2}\Phi_\mu{}^\mu\pm \frac{\text{sh}\gamma}{2}\,\sqrt{2 \Phi_\mu{}^\nu\Phi_\nu{}^\mu-(\Phi_\mu{}^\mu)^2}
    \,,
\end{equation}
with $\Phi_{\mu\nu}=\partial_\mu \phi^i\partial_\nu \phi^i$. 
This is the result found in Ref.~\cite{Conti:2022egv} from dimensional reduction of ModMax, equation~\eqref{eq:MM}. 
Notice also that this flow is non-trivial only for~$N\geq2$, for which $x_1$ and $x_2$ are algebraically independent, otherwise, it is merely a rescaling of the Lagrangian, $L_\gamma=e^{\gamma}L_0$.

\section{Two-Parameter Flow}
Having determined the Root-$\TTbar$ flow, we will now verify that it commutes with the ordinary $\TTbar$ flow. The latter is sourced by the operator~$O$ of Eq.~\eqref{eq:TTbardef}. We can express $O$ in terms of the Lagrangian and its derivatives following what we did for $R$ around Eq.~\eqref{eq:TTminusT2}, and we get the (rather cumbersome) formula
\begin{widetext}
\begin{equation}
\label{eq:TTbar}
    O = - L^2 - 2 \left( \left( \frac{\partial L}{\partial x_1} \right)^2 + 2 \frac{\partial L}{\partial x_1} \frac{\partial L}{\partial x_2} x_1 + 2 \left( \frac{\partial L}{\partial x_2} \right)^2 ( x_1^2 - x_2 ) \right) ( x_1^2 - x_2 ) + 2 L \left( \frac{\partial L}{\partial x_1} x_1 + 2 \frac{\partial L}{\partial x_2} x_2 \right).
\end{equation}
\end{widetext}
Now, consider 
a $\TTbar$ deformation of~$L_{\gamma}$~\eqref{eq:Lgamma}, where the dependence on $x_1$ and $x_2$ appears exclusively through $x_1^{(\gamma)}$ and $x_2^{(\gamma)}$. A change of variables and some algebra, along with the observations~(\ref{eq:constant}--\ref{eq:trace_unchanged}), allows us to reformulate the operator $O$ in terms of $L$ and its partial derivatives \textit{with respect to} $x_1^{(\gamma)}$ and $x_2^{(\gamma)}$. In fact, we find that~$O$ is given precisely by the same expression~\eqref{eq:TTbar} up to promoting $x_1$ and $x_2$ to $x_1^{(\gamma)}$ and $x_2^{(\gamma)}$, respectively.

Therefore, to take a theory defined by $L_{\gamma}$ and $\TTbar$-deform (obtaining $L_{(\lambda,\gamma)}$), we may as well first $\TTbar$-deform $L_0$ (thereby obtaining $L_\lambda$), and then replace everywhere $x_1$ and $x_2$ with $x_1^{(\gamma)}$ and $x_2^{(\gamma)}$. This last step is precisely tantamount to a Root-$\TTbar$ deformation, which goes to show that the diagram of Fig.~\ref{fig:commute} indeed commutes.

The two-parameter Lagrangian can then be found by taking any $\TTbar$-deformed Lagrangian, such as the ones in~\cite{Bonelli:2018kik}, and replacing $x_1$, $x_2$ with the expressions in Eq.~\eqref{eq:X1X2}.
For the special case of $N$ free scalars we find
\begin{equation}\label{N_scalars_modBI}
L_{(\lambda,\gamma)}=\frac{\sqrt{1+2\lambda x_{1}^{(\gamma)}+2\lambda^2\big[\big(x_{1}^{(\gamma)}\big)^2-x_{2}^{(\gamma)}\big]}-1}{2 \lambda}
\,,
\end{equation}
which is precisely the result obtained in Ref.~\cite{Conti:2022egv} from the dimensional reduction of the four-dimensional ModMax-Born-Infeld theory \cite{Bandos:2020jsw,Bandos:2020hgy,Babaei-Aghbolagh:2022uij,Ferko:2022iru}.

\section{Fermions}
It is natural to wonder whether our discussion may be extended to models with Fermions.
It is unclear how to treat a non-analytic function of Grassmann variables only, such as the one that would arise in a Root-$\TTbar$ deformed model involving only Fermions. It is more natural to first look at a theory of Bosons and Fermions, which admits a regular expansion in Grassman variables (with non-analytic coefficients  in the Boson fields).
Let us consider a free model of $N_B$ Bosons and $N_F$ Dirac Fermions,
\begin{equation}
\label{eq:fermionseed}
    L = \frac{1}{2} \partial_\mu \phi^i \partial^\mu \phi^i 
    + \frac{{\rm i}}{2} \psib^I \gamma^\mu \partial_\mu \psi^I 
    - \frac{{\rm i}}{2} \left( \partial_\mu \psib^I  \right)\gamma^\mu \psi^I \, ,
\end{equation}
where
$i=1,\dots N_B$, and
$I=1,\dots N_F$.

To build an ansatz for the $\gamma$-flowed action we need to consider the stress-energy tensor of this model. Here we will consider the Hilbert, rather than Noether, stress-energy tensor. Much like for $\TTbar$ flows~\cite{Baggio:2018rpv}, we expect that the deformed Lagrangian differs in the two choices by terms which vanish on-shell.
It is possible to show that, for theories that depend on the Fermions only via the symmetric tensor $\psib \gamma_{(\mu} \partial_{\nu)} \psi - ( \partial_{(\mu} \psib ) \gamma_{\nu)} \psi$, the Hilbert tensor is the symmetrized Noether tensor. This ensures that the deformed Lagrangian will be a function of
\begin{equation}
\label{eq:fermionX1}
    (X_1)_{\mu\nu}=
     \partial_\mu \phi^i \partial_\nu \phi^i 
     + {\rm i} \psib^I \gamma_{(\mu} \partial_{\nu)} \psi^I 
     - {\rm i}  \big(\partial_{(\mu} \psib^I\big)  \gamma_{\nu)} \psi^I  ,
\end{equation}
and of $(X_2)_{\mu\nu}=(X_1)_{\mu\rho}(X_1)^{\rho}{}_{\nu}$ --- more precisely of their traces $x_1$ and~$x_2$.
Given such a dependence, the flow equation can be determined by truncating all expressions to the Bosons, which immediately reproduces Eq.~\eqref{eq:rotttflow}.

Hence, given an action which depends on~$x_1$ and $x_2$ as in Eq.~\eqref{eq:fermionX1}, the Root-$\TTbar$ flow is found by replacing the $\gamma$-dependent expressions~\eqref{eq:X1X2}.
The same argument of the previous section confirms this flow too commutes with the usual~$\TTbar$ flow.
Note that in \eqref{eq:fermionseed} we started from a system of free Dirac Fermions. However, the same argument concerning the integration of the flows holds if chirality conditions are imposed. 
Interestingly, in the presence of at least one real chiral Fermion the deformed Lagrangian is non-trivial even with~$N_B=1$.

\section{Conclusions and Outlook}
We have discussed  Root-$\TTbar$ deformations, which commute with the $\TTbar$ flow and are  defined in terms of a classically marginal non-analytic operator. One might have expected the deformed theories to be pathological and unnatural. Instead, they are appealing, satisfying a simple flow equation, and encompass the models arising from dimensional reduction of ModMax~\cite{Bandos:2020jsw,Bandos:2020hgy}.
ModMax is also a non-analytic classically conformal theory, with many intriguing properties: it admits various physical solutions such as plane waves~\cite{Bandos:2020jsw} and dyons~\cite{Lechner:2022qhb}, its Hamiltonian is bounded from below~\cite{Escobar:2021mpx}, it can be  supersymmetrized~\cite{Bandos:2021rqy,Kuzenko:2021cvx,Kuzenko:2021qcx}, and it can be related to Maxwell theory coupled to an axion-dilaton~\cite{Lechner:2022qhb} --- see also the recent review~\cite{Sorokin:2021tge}.
While many questions about ModMax remain open, it is certainly a physically interesting theory and it is inspiring that the two-dimensional Root-$\TTbar$ deformation falls in the same class of models.

The main open question is how to quantize such non-analytic 
theories. This will be subtle, but may be possible for Root-$\TTbar$ flows of CFTs, by demanding the deformed theory to be conformal and bootstrapping  its properties. If successful, one could construct new marginal  deformations of generic CFTs. In general, the Root-$\TTbar$ operator will not be part of spectrum of local operators of the theory (with the intriguing exception of the single free Boson). Hence the deformation should result in one-parameter families of theories that are quite different from the familiar local unitary two-dimensional CFTs --- and it would be very interesting to understand such differences. A first step is understanding how $N\geq2$ free Bosons get coupled by the deformation and extending the analysis to WZW models and minimal models where algebraic techniques are very well developed.
The quantization of Root-$\TTbar$-deformed QFTs (as opposed to CFTs) is likely to be harder. By imposing that $\TTbar$ and Root-$\TTbar$ deformations commute, one can first perform the Root-$\TTbar$ deformation (of a CFT) and only later break conformal invariance by~$\TTbar$. This will clarify the general quantum properties of Root-$\TTbar$.

Again with quantization in mind, it would also be useful to study Root-$\TTbar$ in Hamiltonian terms~\cite{Rodriguez:2021tcz,Bagchi:2022nvj}, as was recently done~\cite{Escobar:2021mpx} for~ModMax, and to investigate whether the Root-$\TTbar$ deformation acts in a simple manner on the S-matrix of (integrable and non-integrable) theories, like $\TTbar$ does~\cite{Mussardo:1999aj,Dubovsky:2018bmo}. (See~\cite{Borsato:2022tmu} for recent work on the integrability of Root-$\TTbar$.)
In turn, understanding the quantum properties of Root-$\TTbar$ would also extend our understanding of ModMax, where quantization remains a major stumbling block. This is expected for a $D=4$ model for which no perturbative quantization scheme is available. Approaching this issue from two dimensions, where CFTs are under especially good control, may shed new light on the issue.

Recalling that $\TTbar$ is only one of many current-current deformations~\cite{Smirnov:2016lqw}, all obeying a flow equation~\cite{Hernandez-Chifflet:2019sua}, it would be interesting to see if the higher-spin analogues of~$\TTbar$ also commute with Root-$\TTbar$. Moreover, one could also  consider higher~roots of higher-spin $\TTbar$-like operators. They would also be na\"ively marginal (non-analytic) operators. Their study would probably be difficult, as is the case for higher-spin $\TTbar$ deformations~\cite{Hernandez-Chifflet:2019sua}, but might be instructive already in the case of simple theories, see~\cite{Camilo:2021gro} for~$\TTbar$.
By allowing for the breaking of Lorentz symmetry, more general scale-invariant roots of current-current deformations could also be considered --- for example a  Root-$J\overline{T}$
deformation based on the one of \cite{Guica:2017lia}.

In this letter we have merely initiated the study of models involving Fermions, considering a relatively small class of models. It would be especially interesting to perform a systematic study of the Root-$\TTbar$ deformations of superconformal theories. It is well-established that a supersymmetric model remains supersymmetric under $\TTbar$ flows~\cite{Baggio:2018rpv, Chang:2018dge, Jiang:2019hux,Chang:2019kiu,Coleman:2019dvf}. The fact that ModMax may be supersymmetrized is also encouraging. However, it is not clear that the deformation of a free superconformal action of the type~\eqref{eq:fermionseed} (for example the ${\cal N}=(1,1)$ supersymmetric system when $N_B=N_F$) gives a superconformal model --- this certainly requires the supersymmetry transformations to be modified. In the case of the standard $\TTbar$ deformation it was shown that the operator ${\cal O}$ of Eq.~\eqref{eq:TTbardef} is (on-shell) equivalent to a manifestly supersymmetric supercurrent-squared operator defined in superspace~\cite{Baggio:2018rpv, Chang:2018dge}. It is straightforward to propose manifestly supersymmetric extensions of the Root-$\TTbar$ operator, Eq.~\eqref{Root-TTbar_CFT}, for superconformal field theories. In the simplest case of ${\cal N}=(0,1)$ supersymmetry the following superspace Lagrangian is an example
\begin{equation}
\int d\theta^+ \frac{{\cal S}_{+++}{\cal T}_{----}}{\sqrt{{\cal T}_{++++}{\cal T}_{----}}}
=\sqrt{\TTbar}+\text{Fermions\,}.
\label{Susy-Root-TTbar}
\end{equation}
Here  ${\cal S}_{+++}$ and ${\cal T}_{----}$ define the ${\cal N}=(0,1)$ supercurrents, and 
${\cal T}_{++++}=D_+ {\cal S}_{+++}$, see 
Refs.~\cite{Baggio:2018rpv, Chang:2018dge}.
For a single 
${\cal N}=(0,1)$ real scalar multiplet, the superspace Lagrangian ${\cal A}_-(\gamma)=e^{\gamma}(D_+\Phi)(\partial_{--}\Phi)$
satisfies a classical flow driven by the operator \eqref{Susy-Root-TTbar}.
This is a direct analogue of
the bosonic Root-$\TTbar$ deformation of a single real free scalar field discussed above.
Similar superspace Root-$\TTbar$ operators can be defined for models with more supersymmetry.
Aside from supersymmetry, it would also be interesting to consider the bosonization and fermionization of simple two-dimensional CFTs and their Root-$\TTbar$ flows.

Finally, a major application of $\TTbar$ deformations has been to holography, mapping them to gravitational theories in three-dimensional anti-De Sitter space-time~\cite{McGough:2016lol,Giveon:2017nie}, see also~\cite{Apolo:2019zai,Chakraborty:2020cgo,Coleman:2021nor}.
The holographic interpretation of such a marginal but non-analytic deformation of the boundary CFT$_2$ should be understood more clearly. An immediate application of this letter is to recast the Root-$\TTbar$ deformation in terms of the boundary conditions of the AdS$_3$ metric much like in~\cite{Bzowski:2018pcy}, which will provide a holographic formulation of the deformation and yield a new class of holographic backgrounds~\cite{toappear}.

Holography provided a way to define  new irrelevant deformations in $D>2$ dimensions~\cite{Taylor:2018xcy,Hartman:2018tkw}, in particular in $D=4$ for Maxwell~\cite{Conti:2018jho}, its supersymmetrization~\cite{Ferko:2019oyv} and ModMax theories~\cite{Babaei-Aghbolagh:2022uij,Ferko:2022iru}.
Both our $D=2$ operator $R$~\eqref{eq:rootttbar}, and the $D=4$ operator generating the  $\gamma$-flow of  ModMax-Born-Infeld~\cite{Babaei-Aghbolagh:2022uij} can be written as
\begin{equation}
    \sqrt{-t^{\mu\nu}t_{\mu\nu}}\,,\qquad t_{\mu\nu}=T_{\mu\nu}-\frac{1}{D} g_{\mu\nu}T_\rho{}^\rho\,,
    \label{eq:sqrtTTbar-D}
\end{equation}
where $t_{\mu\nu}$ is the traceless part of $T_{\mu\nu}$. It is natural to wonder whether this operator plays a universal role in generating ModMax-like deformations in any  dimension, and to investigate generic  Root-$\TTbar$-type deformations in $D>2$ models, including holographic ones. One piece of evidence that non-analytic deformations may be relevant for holography in $D>2$ is that operators involving square roots of field gradients have been studied in the context of the fluid-gravity correspondence~\cite{Compere:2011dx} in general dimension~$D$.

A different, but very interesting, direction is the ``generalized holographic principle'' of \cite{Araujo-Regado:2022gvw}, which also involves stress-tensor deformations in $D > 2$; one might wonder whether $\sqrt{TT}$-type operators  may be needed to UV-complete the deformations used in that proposal. Furthermore, the interplay between ultra/non-relativistic limits of a CFT$_2$ and the BMS$_3$ symmetry found in~\cite{Rodriguez:2021tcz,Bagchi:2022nvj} also suggests possible relations between operators like \eqref{eq:sqrtTTbar-D}  and deformations of space-time symmetries in dimensions higher than two. Finally, see also \cite{Garcia:2022wad} for an analogue of the Root-$\TT$ deformation in $(0+1)$ spacetime dimensions.

We believe that addressing all these questions may result in exciting new insights on the structure of conformal and non-conformal quantum field theories and of their holographic duals, and we plan to return to some of these in the near future.

\begin{acknowledgments}
We thank Gleb Arutyunov, Grisha Korchemsky, Dima Sorokin, Marika Taylor, Roberto Volpato, and Aron Wall for helpful discussions, and Dima Sorokin for useful comments on a draft of this letter.

C.\,F. is supported by U.S. Department of Energy grant DE-SC0009999 and by funds from the University of California.
A.\,S. acknowledges support from the European Union -- NextGenerationEU, and from the program STARS@UNIPD, under project ``Exact-Holography'', \textit{A new exact approach to holography: harnessing the power of string
theory, conformal field theory, and integrable models}.
 L.\,S. is supported by a postgraduate
scholarship at the University of Queensland. The work of G.\,T.-M. is supported by the
Australian Research Council (ARC) Future Fellowship FT180100353, and by the Capacity
Building Package of the University of Queensland. 
L.\,S. and G.\,T.-M.
also thank the mathematical research institute MATRIX in Australia where part of this research was performed.

\end{acknowledgments}

\appendix

\bibliographystyle{apsrev4-2} 
\bibliography{apssamp}

\providecommand{\noopsort}[1]{}\providecommand{\singleletter}[1]{#1}%
\begin{thebibliography}{51}%
\makeatletter
\providecommand \@ifxundefined [1]{%
 \@ifx{#1\undefined}
}%
\providecommand \@ifnum [1]{%
 \ifnum #1\expandafter \@firstoftwo
 \else \expandafter \@secondoftwo
 \fi
}%
\providecommand \@ifx [1]{%
 \ifx #1\expandafter \@firstoftwo
 \else \expandafter \@secondoftwo
 \fi
}%
\providecommand \natexlab [1]{#1}%
\providecommand \enquote  [1]{``#1''}%
\providecommand \bibnamefont  [1]{#1}%
\providecommand \bibfnamefont [1]{#1}%
\providecommand \citenamefont [1]{#1}%
\providecommand \href@noop [0]{\@secondoftwo}%
\providecommand \href [0]{\begingroup \@sanitize@url \@href}%
\providecommand \@href[1]{\@@startlink{#1}\@@href}%
\providecommand \@@href[1]{\endgroup#1\@@endlink}%
\providecommand \@sanitize@url [0]{\catcode `\\12\catcode `\$12\catcode
  `\&12\catcode `\#12\catcode `\^12\catcode `\_12\catcode `\%12\relax}%
\providecommand \@@startlink[1]{}%
\providecommand \@@endlink[0]{}%
\providecommand \url  [0]{\begingroup\@sanitize@url \@url }%
\providecommand \@url [1]{\endgroup\@href {#1}{\urlprefix }}%
\providecommand \urlprefix  [0]{URL }%
\providecommand \Eprint [0]{\href }%
\providecommand \doibase [0]{https://doi.org/}%
\providecommand \selectlanguage [0]{\@gobble}%
\providecommand \bibinfo  [0]{\@secondoftwo}%
\providecommand \bibfield  [0]{\@secondoftwo}%
\providecommand \translation [1]{[#1]}%
\providecommand \BibitemOpen [0]{}%
\providecommand \bibitemStop [0]{}%
\providecommand \bibitemNoStop [0]{.\EOS\space}%
\providecommand \EOS [0]{\spacefactor3000\relax}%
\providecommand \BibitemShut  [1]{\csname bibitem#1\endcsname}%
\let\auto@bib@innerbib\@empty
\bibitem [{\citenamefont {Zamolodchikov}(2004)}]{Zamolodchikov:2004ce}%
  \BibitemOpen
  \bibfield  {author} {\bibinfo {author} {\bibfnamefont {A.~B.}\ \bibnamefont
  {Zamolodchikov}},\ }\Eprint {https://arxiv.org/abs/hep-th/0401146}
  {arXiv:hep-th/0401146 [hep-th]}  (\bibinfo {year} {2004})\BibitemShut
  {NoStop}%
\bibitem [{\citenamefont {Smirnov}\ and\ \citenamefont
  {Zamolodchikov}(2017)}]{Smirnov:2016lqw}%
  \BibitemOpen
  \bibfield  {author} {\bibinfo {author} {\bibfnamefont {F.~A.}\ \bibnamefont
  {Smirnov}}\ and\ \bibinfo {author} {\bibfnamefont {A.~B.}\ \bibnamefont
  {Zamolodchikov}},\ }\href {https://doi.org/10.1016/j.nuclphysb.2016.12.014}
  {\bibfield  {journal} {\bibinfo  {journal} {Nucl. Phys. B}\ }\textbf
  {\bibinfo {volume} {915}},\ \bibinfo {pages} {363} (\bibinfo {year}
  {2017})},\ \Eprint {https://arxiv.org/abs/1608.05499} {arXiv:1608.05499
  [hep-th]} \BibitemShut {NoStop}%
\bibitem [{\citenamefont {Cavagli\`a}\ \emph {et~al.}(2016)\citenamefont
  {Cavagli\`a}, \citenamefont {Negro}, \citenamefont {Sz\'ecs\'enyi},\ and\
  \citenamefont {Tateo}}]{Cavaglia:2016oda}%
  \BibitemOpen
  \bibfield  {author} {\bibinfo {author} {\bibfnamefont {A.}~\bibnamefont
  {Cavagli\`a}}, \bibinfo {author} {\bibfnamefont {S.}~\bibnamefont {Negro}},
  \bibinfo {author} {\bibfnamefont {I.~M.}\ \bibnamefont {Sz\'ecs\'enyi}},\
  and\ \bibinfo {author} {\bibfnamefont {R.}~\bibnamefont {Tateo}},\ }\href
  {https://doi.org/10.1007/JHEP10(2016)112} {\bibfield  {journal} {\bibinfo
  {journal} {JHEP}\ }\textbf {\bibinfo {volume} {10}},\ \bibinfo {pages}
  {112}},\ \Eprint {https://arxiv.org/abs/1608.05534} {arXiv:1608.05534
  [hep-th]} \BibitemShut {NoStop}%
\bibitem [{\citenamefont {Baggio}\ and\ \citenamefont
  {Sfondrini}(2018)}]{Baggio:2018gct}%
  \BibitemOpen
  \bibfield  {author} {\bibinfo {author} {\bibfnamefont {M.}~\bibnamefont
  {Baggio}}\ and\ \bibinfo {author} {\bibfnamefont {A.}~\bibnamefont
  {Sfondrini}},\ }\href {https://doi.org/10.1103/PhysRevD.98.021902} {\bibfield
   {journal} {\bibinfo  {journal} {Phys. Rev. D}\ }\textbf {\bibinfo {volume}
  {98}},\ \bibinfo {pages} {021902} (\bibinfo {year} {2018})},\ \Eprint
  {https://arxiv.org/abs/1804.01998} {arXiv:1804.01998 [hep-th]} \BibitemShut
  {NoStop}%
\bibitem [{\citenamefont {Frolov}(2020)}]{Frolov:2019nrr}%
  \BibitemOpen
  \bibfield  {author} {\bibinfo {author} {\bibfnamefont {S.}~\bibnamefont
  {Frolov}},\ }\href {https://doi.org/10.1134/S0081543820030098} {\bibfield
  {journal} {\bibinfo  {journal} {Proc. Steklov Inst. Math.}\ }\textbf
  {\bibinfo {volume} {309}},\ \bibinfo {pages} {107} (\bibinfo {year}
  {2020})},\ \Eprint {https://arxiv.org/abs/1905.07946} {arXiv:1905.07946
  [hep-th]} \BibitemShut {NoStop}%
\bibitem [{\citenamefont {McGough}\ \emph {et~al.}(2018)\citenamefont
  {McGough}, \citenamefont {Mezei},\ and\ \citenamefont
  {Verlinde}}]{McGough:2016lol}%
  \BibitemOpen
  \bibfield  {author} {\bibinfo {author} {\bibfnamefont {L.}~\bibnamefont
  {McGough}}, \bibinfo {author} {\bibfnamefont {M.}~\bibnamefont {Mezei}},\
  and\ \bibinfo {author} {\bibfnamefont {H.}~\bibnamefont {Verlinde}},\ }\href
  {https://doi.org/10.1007/JHEP04(2018)010} {\bibfield  {journal} {\bibinfo
  {journal} {JHEP}\ }\textbf {\bibinfo {volume} {04}},\ \bibinfo {pages}
  {010}},\ \Eprint {https://arxiv.org/abs/1611.03470} {arXiv:1611.03470
  [hep-th]} \BibitemShut {NoStop}%
\bibitem [{\citenamefont {Giveon}\ \emph {et~al.}(2017)\citenamefont {Giveon},
  \citenamefont {Itzhaki},\ and\ \citenamefont {Kutasov}}]{Giveon:2017nie}%
  \BibitemOpen
  \bibfield  {author} {\bibinfo {author} {\bibfnamefont {A.}~\bibnamefont
  {Giveon}}, \bibinfo {author} {\bibfnamefont {N.}~\bibnamefont {Itzhaki}},\
  and\ \bibinfo {author} {\bibfnamefont {D.}~\bibnamefont {Kutasov}},\ }\href
  {https://doi.org/10.1007/JHEP07(2017)122} {\bibfield  {journal} {\bibinfo
  {journal} {JHEP}\ }\textbf {\bibinfo {volume} {07}},\ \bibinfo {pages}
  {122}},\ \Eprint {https://arxiv.org/abs/1701.05576} {arXiv:1701.05576
  [hep-th]} \BibitemShut {NoStop}%
\bibitem [{\citenamefont {Dubovsky}\ \emph {et~al.}(2017)\citenamefont
  {Dubovsky}, \citenamefont {Gorbenko},\ and\ \citenamefont
  {Mirbabayi}}]{Dubovsky:2017cnj}%
  \BibitemOpen
  \bibfield  {author} {\bibinfo {author} {\bibfnamefont {S.}~\bibnamefont
  {Dubovsky}}, \bibinfo {author} {\bibfnamefont {V.}~\bibnamefont {Gorbenko}},\
  and\ \bibinfo {author} {\bibfnamefont {M.}~\bibnamefont {Mirbabayi}},\ }\href
  {https://doi.org/10.1007/JHEP09(2017)136} {\bibfield  {journal} {\bibinfo
  {journal} {JHEP}\ }\textbf {\bibinfo {volume} {09}},\ \bibinfo {pages}
  {136}},\ \Eprint {https://arxiv.org/abs/1706.06604} {arXiv:1706.06604
  [hep-th]} \BibitemShut {NoStop}%
\bibitem [{\citenamefont {Dubovsky}\ \emph {et~al.}(2018)\citenamefont
  {Dubovsky}, \citenamefont {Gorbenko},\ and\ \citenamefont
  {Hern\'andez-Chifflet}}]{Dubovsky:2018bmo}%
  \BibitemOpen
  \bibfield  {author} {\bibinfo {author} {\bibfnamefont {S.}~\bibnamefont
  {Dubovsky}}, \bibinfo {author} {\bibfnamefont {V.}~\bibnamefont {Gorbenko}},\
  and\ \bibinfo {author} {\bibfnamefont {G.}~\bibnamefont
  {Hern\'andez-Chifflet}},\ }\href {https://doi.org/10.1007/JHEP09(2018)158}
  {\bibfield  {journal} {\bibinfo  {journal} {JHEP}\ }\textbf {\bibinfo
  {volume} {09}},\ \bibinfo {pages} {158}},\ \Eprint
  {https://arxiv.org/abs/1805.07386} {arXiv:1805.07386 [hep-th]} \BibitemShut
  {NoStop}%
\bibitem [{\citenamefont {Conti}\ \emph {et~al.}(2019)\citenamefont {Conti},
  \citenamefont {Negro},\ and\ \citenamefont {Tateo}}]{Conti:2018tca}%
  \BibitemOpen
  \bibfield  {author} {\bibinfo {author} {\bibfnamefont {R.}~\bibnamefont
  {Conti}}, \bibinfo {author} {\bibfnamefont {S.}~\bibnamefont {Negro}},\ and\
  \bibinfo {author} {\bibfnamefont {R.}~\bibnamefont {Tateo}},\ }\href
  {https://doi.org/10.1007/JHEP02(2019)085} {\bibfield  {journal} {\bibinfo
  {journal} {JHEP}\ }\textbf {\bibinfo {volume} {02}},\ \bibinfo {pages}
  {085}},\ \Eprint {https://arxiv.org/abs/1809.09593} {arXiv:1809.09593
  [hep-th]} \BibitemShut {NoStop}%
\bibitem [{\citenamefont {Cardy}(2018)}]{Cardy:2018sdv}%
  \BibitemOpen
  \bibfield  {author} {\bibinfo {author} {\bibfnamefont {J.}~\bibnamefont
  {Cardy}},\ }\href {https://doi.org/10.1007/JHEP10(2018)186} {\bibfield
  {journal} {\bibinfo  {journal} {JHEP}\ }\textbf {\bibinfo {volume} {10}},\
  \bibinfo {pages} {186}},\ \Eprint {https://arxiv.org/abs/1801.06895}
  {arXiv:1801.06895 [hep-th]} \BibitemShut {NoStop}%
\bibitem [{\citenamefont {Jiang}(2021)}]{Jiang:2019epa}%
  \BibitemOpen
  \bibfield  {author} {\bibinfo {author} {\bibfnamefont {Y.}~\bibnamefont
  {Jiang}},\ }\href {https://doi.org/10.1088/1572-9494/abe4c9} {\bibfield
  {journal} {\bibinfo  {journal} {Commun. Theor. Phys.}\ }\textbf {\bibinfo
  {volume} {73}},\ \bibinfo {pages} {057201} (\bibinfo {year} {2021})},\
  \Eprint {https://arxiv.org/abs/1904.13376} {arXiv:1904.13376 [hep-th]}
  \BibitemShut {NoStop}%
\bibitem [{\citenamefont {Ferko}(2021)}]{Ferko:2021loo}%
  \BibitemOpen
  \bibfield  {author} {\bibinfo {author} {\bibfnamefont {C.}~\bibnamefont
  {Ferko}},\ }\emph {\bibinfo {title} {{Supersymmetry and Irrelevant
  Deformations}}},\ \href@noop {} {Ph.D. thesis},\ \bibinfo  {school} {Chicago
  U.} (\bibinfo {year} {2021}),\ \Eprint {https://arxiv.org/abs/2112.14647}
  {arXiv:2112.14647 [hep-th]} \BibitemShut {NoStop}%
\bibitem [{\citenamefont {Rodr\'\i{}guez}\ \emph {et~al.}(2021)\citenamefont
  {Rodr\'\i{}guez}, \citenamefont {Tempo},\ and\ \citenamefont
  {Troncoso}}]{Rodriguez:2021tcz}%
  \BibitemOpen
  \bibfield  {author} {\bibinfo {author} {\bibfnamefont {P.}~\bibnamefont
  {Rodr\'\i{}guez}}, \bibinfo {author} {\bibfnamefont {D.}~\bibnamefont
  {Tempo}},\ and\ \bibinfo {author} {\bibfnamefont {R.}~\bibnamefont
  {Troncoso}},\ }\href {https://doi.org/10.1007/JHEP11(2021)133} {\bibfield
  {journal} {\bibinfo  {journal} {JHEP}\ }\textbf {\bibinfo {volume} {11}},\
  \bibinfo {pages} {133}},\ \Eprint {https://arxiv.org/abs/2106.09750}
  {arXiv:2106.09750 [hep-th]} \BibitemShut {NoStop}%
\bibitem [{\citenamefont {Bagchi}\ \emph {et~al.}(2022)\citenamefont {Bagchi},
  \citenamefont {Banerjee},\ and\ \citenamefont {Muraki}}]{Bagchi:2022nvj}%
  \BibitemOpen
  \bibfield  {author} {\bibinfo {author} {\bibfnamefont {A.}~\bibnamefont
  {Bagchi}}, \bibinfo {author} {\bibfnamefont {A.}~\bibnamefont {Banerjee}},\
  and\ \bibinfo {author} {\bibfnamefont {H.}~\bibnamefont {Muraki}},\ }\Eprint
  {https://arxiv.org/abs/2205.05094} {arXiv:2205.05094 [hep-th]}  (\bibinfo
  {year} {2022})\BibitemShut {NoStop}%
\bibitem [{\citenamefont {Bandos}\ \emph {et~al.}(2020)\citenamefont {Bandos},
  \citenamefont {Lechner}, \citenamefont {Sorokin},\ and\ \citenamefont
  {Townsend}}]{Bandos:2020jsw}%
  \BibitemOpen
  \bibfield  {author} {\bibinfo {author} {\bibfnamefont {I.}~\bibnamefont
  {Bandos}}, \bibinfo {author} {\bibfnamefont {K.}~\bibnamefont {Lechner}},
  \bibinfo {author} {\bibfnamefont {D.}~\bibnamefont {Sorokin}},\ and\ \bibinfo
  {author} {\bibfnamefont {P.~K.}\ \bibnamefont {Townsend}},\ }\href
  {https://doi.org/10.1103/PhysRevD.102.121703} {\bibfield  {journal} {\bibinfo
   {journal} {Phys. Rev. D}\ }\textbf {\bibinfo {volume} {102}},\ \bibinfo
  {pages} {121703} (\bibinfo {year} {2020})},\ \Eprint
  {https://arxiv.org/abs/2007.09092} {arXiv:2007.09092 [hep-th]} \BibitemShut
  {NoStop}%
\bibitem [{\citenamefont {Bandos}\ \emph
  {et~al.}(2021{\natexlab{a}})\citenamefont {Bandos}, \citenamefont {Lechner},
  \citenamefont {Sorokin},\ and\ \citenamefont {Townsend}}]{Bandos:2020hgy}%
  \BibitemOpen
  \bibfield  {author} {\bibinfo {author} {\bibfnamefont {I.}~\bibnamefont
  {Bandos}}, \bibinfo {author} {\bibfnamefont {K.}~\bibnamefont {Lechner}},
  \bibinfo {author} {\bibfnamefont {D.}~\bibnamefont {Sorokin}},\ and\ \bibinfo
  {author} {\bibfnamefont {P.~K.}\ \bibnamefont {Townsend}},\ }\href
  {https://doi.org/10.1007/JHEP03(2021)022} {\bibfield  {journal} {\bibinfo
  {journal} {JHEP}\ }\textbf {\bibinfo {volume} {03}},\ \bibinfo {pages}
  {022}},\ \Eprint {https://arxiv.org/abs/2012.09286} {arXiv:2012.09286
  [hep-th]} \BibitemShut {NoStop}%
\bibitem [{\citenamefont {Kosyakov}(2020)}]{Kosyakov:2020wxv}%
  \BibitemOpen
  \bibfield  {author} {\bibinfo {author} {\bibfnamefont {B.~P.}\ \bibnamefont
  {Kosyakov}},\ }\href {https://doi.org/10.1016/j.physletb.2020.135840}
  {\bibfield  {journal} {\bibinfo  {journal} {Phys. Lett. B}\ }\textbf
  {\bibinfo {volume} {810}},\ \bibinfo {pages} {135840} (\bibinfo {year}
  {2020})},\ \Eprint {https://arxiv.org/abs/2007.13878} {arXiv:2007.13878
  [hep-th]} \BibitemShut {NoStop}%
\bibitem [{\citenamefont {Sorokin}(2021)}]{Sorokin:2021tge}%
  \BibitemOpen
  \bibfield  {author} {\bibinfo {author} {\bibfnamefont {D.~P.}\ \bibnamefont
  {Sorokin}}\ }(\bibinfo {year} {2021})\ \Eprint
  {https://arxiv.org/abs/2112.12118} {arXiv:2112.12118 [hep-th]} \BibitemShut
  {NoStop}%
\bibitem [{\citenamefont {Conti}\ \emph {et~al.}(2018)\citenamefont {Conti},
  \citenamefont {Iannella}, \citenamefont {Negro},\ and\ \citenamefont
  {Tateo}}]{Conti:2018jho}%
  \BibitemOpen
  \bibfield  {author} {\bibinfo {author} {\bibfnamefont {R.}~\bibnamefont
  {Conti}}, \bibinfo {author} {\bibfnamefont {L.}~\bibnamefont {Iannella}},
  \bibinfo {author} {\bibfnamefont {S.}~\bibnamefont {Negro}},\ and\ \bibinfo
  {author} {\bibfnamefont {R.}~\bibnamefont {Tateo}},\ }\href
  {https://doi.org/10.1007/JHEP11(2018)007} {\bibfield  {journal} {\bibinfo
  {journal} {JHEP}\ }\textbf {\bibinfo {volume} {11}},\ \bibinfo {pages}
  {007}},\ \Eprint {https://arxiv.org/abs/1806.11515} {arXiv:1806.11515
  [hep-th]} \BibitemShut {NoStop}%
\bibitem [{\citenamefont {Babaei-Aghbolagh}\ \emph
  {et~al.}(2022{\natexlab{a}})\citenamefont {Babaei-Aghbolagh}, \citenamefont
  {Velni}, \citenamefont {Yekta},\ and\ \citenamefont
  {Mohammadzadeh}}]{Babaei-Aghbolagh:2022uij}%
  \BibitemOpen
  \bibfield  {author} {\bibinfo {author} {\bibfnamefont {H.}~\bibnamefont
  {Babaei-Aghbolagh}}, \bibinfo {author} {\bibfnamefont {K.~B.}\ \bibnamefont
  {Velni}}, \bibinfo {author} {\bibfnamefont {D.~M.}\ \bibnamefont {Yekta}},\
  and\ \bibinfo {author} {\bibfnamefont {H.}~\bibnamefont {Mohammadzadeh}},\
  }\href {https://doi.org/10.1016/j.physletb.2022.137079} {\bibfield  {journal}
  {\bibinfo  {journal} {Phys. Lett. B}\ }\textbf {\bibinfo {volume} {829}},\
  \bibinfo {pages} {137079} (\bibinfo {year} {2022}{\natexlab{a}})},\ \Eprint
  {https://arxiv.org/abs/2202.11156} {arXiv:2202.11156 [hep-th]} \BibitemShut
  {NoStop}%
\bibitem [{\citenamefont {Ferko}\ \emph {et~al.}(2022)\citenamefont {Ferko},
  \citenamefont {Smith},\ and\ \citenamefont
  {Tartaglino-Mazzucchelli}}]{Ferko:2022iru}%
  \BibitemOpen
  \bibfield  {author} {\bibinfo {author} {\bibfnamefont {C.}~\bibnamefont
  {Ferko}}, \bibinfo {author} {\bibfnamefont {L.}~\bibnamefont {Smith}},\ and\
  \bibinfo {author} {\bibfnamefont {G.}~\bibnamefont
  {Tartaglino-Mazzucchelli}},\ }\Eprint {https://arxiv.org/abs/2203.01085}
  {arXiv:2203.01085 [hep-th]}  (\bibinfo {year} {2022})\BibitemShut {NoStop}%
\bibitem [{\citenamefont {Conti}\ \emph {et~al.}(2022)\citenamefont {Conti},
  \citenamefont {Romano},\ and\ \citenamefont {Tateo}}]{Conti:2022egv}%
  \BibitemOpen
  \bibfield  {author} {\bibinfo {author} {\bibfnamefont {R.}~\bibnamefont
  {Conti}}, \bibinfo {author} {\bibfnamefont {J.}~\bibnamefont {Romano}},\ and\
  \bibinfo {author} {\bibfnamefont {R.}~\bibnamefont {Tateo}},\ }\Eprint
  {https://arxiv.org/abs/2206.03415} {arXiv:2206.03415 [hep-th]}  (\bibinfo
  {year} {2022})\BibitemShut {NoStop}%
\bibitem [{\citenamefont {Babaei-Aghbolagh}\ \emph
  {et~al.}(2022{\natexlab{b}})\citenamefont {Babaei-Aghbolagh}, \citenamefont
  {Babaei~Velni}, \citenamefont {Yekta},\ and\ \citenamefont
  {Mohammadzadeh}}]{Babaei-Aghbolagh:2022kfz}%
  \BibitemOpen
  \bibfield  {author} {\bibinfo {author} {\bibfnamefont {H.}~\bibnamefont
  {Babaei-Aghbolagh}}, \bibinfo {author} {\bibfnamefont {K.}~\bibnamefont
  {Babaei~Velni}}, \bibinfo {author} {\bibfnamefont {D.~M.}\ \bibnamefont
  {Yekta}},\ and\ \bibinfo {author} {\bibfnamefont {H.}~\bibnamefont
  {Mohammadzadeh}},\ }\href@noop {} {\  (\bibinfo {year}
  {2022}{\natexlab{b}})},\ \Eprint {https://arxiv.org/abs/2206.12677}
  {arXiv:2206.12677 [hep-th]} \BibitemShut {NoStop}%
\bibitem [{\citenamefont {Bonelli}\ \emph {et~al.}(2018)\citenamefont
  {Bonelli}, \citenamefont {Doroud},\ and\ \citenamefont
  {Zhu}}]{Bonelli:2018kik}%
  \BibitemOpen
  \bibfield  {author} {\bibinfo {author} {\bibfnamefont {G.}~\bibnamefont
  {Bonelli}}, \bibinfo {author} {\bibfnamefont {N.}~\bibnamefont {Doroud}},\
  and\ \bibinfo {author} {\bibfnamefont {M.}~\bibnamefont {Zhu}},\ }\href
  {https://doi.org/10.1007/JHEP06(2018)149} {\bibfield  {journal} {\bibinfo
  {journal} {JHEP}\ }\textbf {\bibinfo {volume} {06}},\ \bibinfo {pages}
  {149}},\ \Eprint {https://arxiv.org/abs/1804.10967} {arXiv:1804.10967
  [hep-th]} \BibitemShut {NoStop}%
\bibitem [{\citenamefont {Baggio}\ \emph {et~al.}(2019)\citenamefont {Baggio},
  \citenamefont {Sfondrini}, \citenamefont {Tartaglino-Mazzucchelli},\ and\
  \citenamefont {Walsh}}]{Baggio:2018rpv}%
  \BibitemOpen
  \bibfield  {author} {\bibinfo {author} {\bibfnamefont {M.}~\bibnamefont
  {Baggio}}, \bibinfo {author} {\bibfnamefont {A.}~\bibnamefont {Sfondrini}},
  \bibinfo {author} {\bibfnamefont {G.}~\bibnamefont
  {Tartaglino-Mazzucchelli}},\ and\ \bibinfo {author} {\bibfnamefont
  {H.}~\bibnamefont {Walsh}},\ }\href {https://doi.org/10.1007/JHEP06(2019)063}
  {\bibfield  {journal} {\bibinfo  {journal} {JHEP}\ }\textbf {\bibinfo
  {volume} {06}},\ \bibinfo {pages} {063}},\ \Eprint
  {https://arxiv.org/abs/1811.00533} {arXiv:1811.00533 [hep-th]} \BibitemShut
  {NoStop}%
\bibitem [{\citenamefont {Lechner}\ \emph {et~al.}(2022)\citenamefont
  {Lechner}, \citenamefont {Marchetti}, \citenamefont {Sainaghi},\ and\
  \citenamefont {Sorokin}}]{Lechner:2022qhb}%
  \BibitemOpen
  \bibfield  {author} {\bibinfo {author} {\bibfnamefont {K.}~\bibnamefont
  {Lechner}}, \bibinfo {author} {\bibfnamefont {P.}~\bibnamefont {Marchetti}},
  \bibinfo {author} {\bibfnamefont {A.}~\bibnamefont {Sainaghi}},\ and\
  \bibinfo {author} {\bibfnamefont {D.~P.}\ \bibnamefont {Sorokin}},\ }\Eprint
  {https://arxiv.org/abs/2206.04657} {arXiv:2206.04657 [hep-th]}  (\bibinfo
  {year} {2022})\BibitemShut {NoStop}%
\bibitem [{\citenamefont {Escobar}\ \emph {et~al.}(2022)\citenamefont
  {Escobar}, \citenamefont {Linares},\ and\ \citenamefont
  {Tlatelpa-Mascote}}]{Escobar:2021mpx}%
  \BibitemOpen
  \bibfield  {author} {\bibinfo {author} {\bibfnamefont {C.~A.}\ \bibnamefont
  {Escobar}}, \bibinfo {author} {\bibfnamefont {R.}~\bibnamefont {Linares}},\
  and\ \bibinfo {author} {\bibfnamefont {B.}~\bibnamefont {Tlatelpa-Mascote}},\
  }\href {https://doi.org/10.1142/S0217751X22500117} {\bibfield  {journal}
  {\bibinfo  {journal} {Int. J. Mod. Phys. A}\ }\textbf {\bibinfo {volume}
  {37}},\ \bibinfo {pages} {2250011} (\bibinfo {year} {2022})},\ \Eprint
  {https://arxiv.org/abs/2112.10060} {arXiv:2112.10060 [hep-th]} \BibitemShut
  {NoStop}%
\bibitem [{\citenamefont {Bandos}\ \emph
  {et~al.}(2021{\natexlab{b}})\citenamefont {Bandos}, \citenamefont {Lechner},
  \citenamefont {Sorokin},\ and\ \citenamefont {Townsend}}]{Bandos:2021rqy}%
  \BibitemOpen
  \bibfield  {author} {\bibinfo {author} {\bibfnamefont {I.}~\bibnamefont
  {Bandos}}, \bibinfo {author} {\bibfnamefont {K.}~\bibnamefont {Lechner}},
  \bibinfo {author} {\bibfnamefont {D.}~\bibnamefont {Sorokin}},\ and\ \bibinfo
  {author} {\bibfnamefont {P.~K.}\ \bibnamefont {Townsend}},\ }\href
  {https://doi.org/10.1007/JHEP10(2021)031} {\bibfield  {journal} {\bibinfo
  {journal} {JHEP}\ }\textbf {\bibinfo {volume} {10}},\ \bibinfo {pages}
  {031}},\ \Eprint {https://arxiv.org/abs/2106.07547} {arXiv:2106.07547
  [hep-th]} \BibitemShut {NoStop}%
\bibitem [{\citenamefont {Kuzenko}(2021)}]{Kuzenko:2021cvx}%
  \BibitemOpen
  \bibfield  {author} {\bibinfo {author} {\bibfnamefont {S.~M.}\ \bibnamefont
  {Kuzenko}},\ }\href {https://doi.org/10.1007/JHEP09(2021)180} {\bibfield
  {journal} {\bibinfo  {journal} {JHEP}\ }\textbf {\bibinfo {volume} {09}},\
  \bibinfo {pages} {180}},\ \Eprint {https://arxiv.org/abs/2106.07173}
  {arXiv:2106.07173 [hep-th]} \BibitemShut {NoStop}%
\bibitem [{\citenamefont {Kuzenko}\ and\ \citenamefont
  {Raptakis}(2021)}]{Kuzenko:2021qcx}%
  \BibitemOpen
  \bibfield  {author} {\bibinfo {author} {\bibfnamefont {S.~M.}\ \bibnamefont
  {Kuzenko}}\ and\ \bibinfo {author} {\bibfnamefont {E.~S.~N.}\ \bibnamefont
  {Raptakis}},\ }\href {https://doi.org/10.1103/PhysRevD.104.125003} {\bibfield
   {journal} {\bibinfo  {journal} {Phys. Rev. D}\ }\textbf {\bibinfo {volume}
  {104}},\ \bibinfo {pages} {125003} (\bibinfo {year} {2021})},\ \Eprint
  {https://arxiv.org/abs/2107.02001} {arXiv:2107.02001 [hep-th]} \BibitemShut
  {NoStop}%
\bibitem [{\citenamefont {Mussardo}\ and\ \citenamefont
  {Simon}(2000)}]{Mussardo:1999aj}%
  \BibitemOpen
  \bibfield  {author} {\bibinfo {author} {\bibfnamefont {G.}~\bibnamefont
  {Mussardo}}\ and\ \bibinfo {author} {\bibfnamefont {P.}~\bibnamefont
  {Simon}},\ }\href {https://doi.org/10.1016/S0550-3213(99)00806-8} {\bibfield
  {journal} {\bibinfo  {journal} {Nucl. Phys. B}\ }\textbf {\bibinfo {volume}
  {578}},\ \bibinfo {pages} {527} (\bibinfo {year} {2000})},\ \Eprint
  {https://arxiv.org/abs/hep-th/9903072} {arXiv:hep-th/9903072} \BibitemShut
  {NoStop}%
\bibitem [{\citenamefont {Borsato}\ \emph {et~al.}(2022)\citenamefont
  {Borsato}, \citenamefont {Ferko},\ and\ \citenamefont
  {Sfondrini}}]{Borsato:2022tmu}%
  \BibitemOpen
  \bibfield  {author} {\bibinfo {author} {\bibfnamefont {R.}~\bibnamefont
  {Borsato}}, \bibinfo {author} {\bibfnamefont {C.}~\bibnamefont {Ferko}},\
  and\ \bibinfo {author} {\bibfnamefont {A.}~\bibnamefont {Sfondrini}},\
  }\href@noop {} {\  (\bibinfo {year} {2022})},\ \Eprint
  {https://arxiv.org/abs/2209.14274} {arXiv:2209.14274 [hep-th]} \BibitemShut
  {NoStop}%
\bibitem [{\citenamefont {Hern\'andez-Chifflet}\ \emph
  {et~al.}(2020)\citenamefont {Hern\'andez-Chifflet}, \citenamefont {Negro},\
  and\ \citenamefont {Sfondrini}}]{Hernandez-Chifflet:2019sua}%
  \BibitemOpen
  \bibfield  {author} {\bibinfo {author} {\bibfnamefont {G.}~\bibnamefont
  {Hern\'andez-Chifflet}}, \bibinfo {author} {\bibfnamefont {S.}~\bibnamefont
  {Negro}},\ and\ \bibinfo {author} {\bibfnamefont {A.}~\bibnamefont
  {Sfondrini}},\ }\href {https://doi.org/10.1103/PhysRevLett.124.200601}
  {\bibfield  {journal} {\bibinfo  {journal} {Phys. Rev. Lett.}\ }\textbf
  {\bibinfo {volume} {124}},\ \bibinfo {pages} {200601} (\bibinfo {year}
  {2020})},\ \Eprint {https://arxiv.org/abs/1911.12233} {arXiv:1911.12233
  [hep-th]} \BibitemShut {NoStop}%
\bibitem [{\citenamefont {Camilo}\ \emph {et~al.}(2021)\citenamefont {Camilo},
  \citenamefont {Fleury}, \citenamefont {Lencs\'es}, \citenamefont {Negro},\
  and\ \citenamefont {Zamolodchikov}}]{Camilo:2021gro}%
  \BibitemOpen
  \bibfield  {author} {\bibinfo {author} {\bibfnamefont {G.}~\bibnamefont
  {Camilo}}, \bibinfo {author} {\bibfnamefont {T.}~\bibnamefont {Fleury}},
  \bibinfo {author} {\bibfnamefont {M.}~\bibnamefont {Lencs\'es}}, \bibinfo
  {author} {\bibfnamefont {S.}~\bibnamefont {Negro}},\ and\ \bibinfo {author}
  {\bibfnamefont {A.}~\bibnamefont {Zamolodchikov}},\ }\href
  {https://doi.org/10.1007/JHEP10(2021)062} {\bibfield  {journal} {\bibinfo
  {journal} {JHEP}\ }\textbf {\bibinfo {volume} {10}},\ \bibinfo {pages}
  {062}},\ \Eprint {https://arxiv.org/abs/2106.11999} {arXiv:2106.11999
  [hep-th]} \BibitemShut {NoStop}%
\bibitem [{\citenamefont {Guica}(2018)}]{Guica:2017lia}%
  \BibitemOpen
  \bibfield  {author} {\bibinfo {author} {\bibfnamefont {M.}~\bibnamefont
  {Guica}},\ }\href {https://doi.org/10.21468/SciPostPhys.5.5.048} {\bibfield
  {journal} {\bibinfo  {journal} {SciPost Phys.}\ }\textbf {\bibinfo {volume}
  {5}},\ \bibinfo {pages} {048} (\bibinfo {year} {2018})},\ \Eprint
  {https://arxiv.org/abs/1710.08415} {arXiv:1710.08415 [hep-th]} \BibitemShut
  {NoStop}%
\bibitem [{\citenamefont {Chang}\ \emph {et~al.}(2019)\citenamefont {Chang},
  \citenamefont {Ferko},\ and\ \citenamefont {Sethi}}]{Chang:2018dge}%
  \BibitemOpen
  \bibfield  {author} {\bibinfo {author} {\bibfnamefont {C.-K.}\ \bibnamefont
  {Chang}}, \bibinfo {author} {\bibfnamefont {C.}~\bibnamefont {Ferko}},\ and\
  \bibinfo {author} {\bibfnamefont {S.}~\bibnamefont {Sethi}},\ }\href
  {https://doi.org/10.1007/JHEP04(2019)131} {\bibfield  {journal} {\bibinfo
  {journal} {JHEP}\ }\textbf {\bibinfo {volume} {04}},\ \bibinfo {pages}
  {131}},\ \Eprint {https://arxiv.org/abs/1811.01895} {arXiv:1811.01895
  [hep-th]} \BibitemShut {NoStop}%
\bibitem [{\citenamefont {Jiang}\ \emph {et~al.}(2019)\citenamefont {Jiang},
  \citenamefont {Sfondrini},\ and\ \citenamefont
  {Tartaglino-Mazzucchelli}}]{Jiang:2019hux}%
  \BibitemOpen
  \bibfield  {author} {\bibinfo {author} {\bibfnamefont {H.}~\bibnamefont
  {Jiang}}, \bibinfo {author} {\bibfnamefont {A.}~\bibnamefont {Sfondrini}},\
  and\ \bibinfo {author} {\bibfnamefont {G.}~\bibnamefont
  {Tartaglino-Mazzucchelli}},\ }\href
  {https://doi.org/10.1103/PhysRevD.100.046017} {\bibfield  {journal} {\bibinfo
   {journal} {Phys. Rev. D}\ }\textbf {\bibinfo {volume} {100}},\ \bibinfo
  {pages} {046017} (\bibinfo {year} {2019})},\ \Eprint
  {https://arxiv.org/abs/1904.04760} {arXiv:1904.04760 [hep-th]} \BibitemShut
  {NoStop}%
\bibitem [{\citenamefont {Chang}\ \emph {et~al.}(2020)\citenamefont {Chang},
  \citenamefont {Ferko}, \citenamefont {Sethi}, \citenamefont {Sfondrini},\
  and\ \citenamefont {Tartaglino-Mazzucchelli}}]{Chang:2019kiu}%
  \BibitemOpen
  \bibfield  {author} {\bibinfo {author} {\bibfnamefont {C.-K.}\ \bibnamefont
  {Chang}}, \bibinfo {author} {\bibfnamefont {C.}~\bibnamefont {Ferko}},
  \bibinfo {author} {\bibfnamefont {S.}~\bibnamefont {Sethi}}, \bibinfo
  {author} {\bibfnamefont {A.}~\bibnamefont {Sfondrini}},\ and\ \bibinfo
  {author} {\bibfnamefont {G.}~\bibnamefont {Tartaglino-Mazzucchelli}},\ }\href
  {https://doi.org/10.1103/PhysRevD.101.026008} {\bibfield  {journal} {\bibinfo
   {journal} {Phys. Rev. D}\ }\textbf {\bibinfo {volume} {101}},\ \bibinfo
  {pages} {026008} (\bibinfo {year} {2020})},\ \Eprint
  {https://arxiv.org/abs/1906.00467} {arXiv:1906.00467 [hep-th]} \BibitemShut
  {NoStop}%
\bibitem [{\citenamefont {Coleman}\ \emph {et~al.}(2019)\citenamefont
  {Coleman}, \citenamefont {Aguilera-Damia}, \citenamefont {Freedman},\ and\
  \citenamefont {Soni}}]{Coleman:2019dvf}%
  \BibitemOpen
  \bibfield  {author} {\bibinfo {author} {\bibfnamefont {E.~A.}\ \bibnamefont
  {Coleman}}, \bibinfo {author} {\bibfnamefont {J.}~\bibnamefont
  {Aguilera-Damia}}, \bibinfo {author} {\bibfnamefont {D.~Z.}\ \bibnamefont
  {Freedman}},\ and\ \bibinfo {author} {\bibfnamefont {R.~M.}\ \bibnamefont
  {Soni}},\ }\href {https://doi.org/10.1007/JHEP10(2019)080} {\bibfield
  {journal} {\bibinfo  {journal} {JHEP}\ }\textbf {\bibinfo {volume} {10}},\
  \bibinfo {pages} {080}},\ \Eprint {https://arxiv.org/abs/1906.05439}
  {arXiv:1906.05439 [hep-th]} \BibitemShut {NoStop}%
\bibitem [{\citenamefont {Apolo}\ \emph {et~al.}(2020)\citenamefont {Apolo},
  \citenamefont {Detournay},\ and\ \citenamefont {Song}}]{Apolo:2019zai}%
  \BibitemOpen
  \bibfield  {author} {\bibinfo {author} {\bibfnamefont {L.}~\bibnamefont
  {Apolo}}, \bibinfo {author} {\bibfnamefont {S.}~\bibnamefont {Detournay}},\
  and\ \bibinfo {author} {\bibfnamefont {W.}~\bibnamefont {Song}},\ }\href
  {https://doi.org/10.1007/JHEP06(2020)109} {\bibfield  {journal} {\bibinfo
  {journal} {JHEP}\ }\textbf {\bibinfo {volume} {06}},\ \bibinfo {pages}
  {109}},\ \Eprint {https://arxiv.org/abs/1911.12359} {arXiv:1911.12359
  [hep-th]} \BibitemShut {NoStop}%
\bibitem [{\citenamefont {Chakraborty}\ \emph {et~al.}(2020)\citenamefont
  {Chakraborty}, \citenamefont {Giveon},\ and\ \citenamefont
  {Kutasov}}]{Chakraborty:2020cgo}%
  \BibitemOpen
  \bibfield  {author} {\bibinfo {author} {\bibfnamefont {S.}~\bibnamefont
  {Chakraborty}}, \bibinfo {author} {\bibfnamefont {A.}~\bibnamefont
  {Giveon}},\ and\ \bibinfo {author} {\bibfnamefont {D.}~\bibnamefont
  {Kutasov}},\ }\href {https://doi.org/10.1007/JHEP11(2020)057} {\bibfield
  {journal} {\bibinfo  {journal} {JHEP}\ }\textbf {\bibinfo {volume} {11}},\
  \bibinfo {pages} {057}},\ \Eprint {https://arxiv.org/abs/2009.03929}
  {arXiv:2009.03929 [hep-th]} \BibitemShut {NoStop}%
\bibitem [{\citenamefont {Coleman}\ \emph {et~al.}(2021)\citenamefont
  {Coleman}, \citenamefont {Mazenc}, \citenamefont {Shyam}, \citenamefont
  {Silverstein}, \citenamefont {Soni}, \citenamefont {Torroba},\ and\
  \citenamefont {Yang}}]{Coleman:2021nor}%
  \BibitemOpen
  \bibfield  {author} {\bibinfo {author} {\bibfnamefont {E.}~\bibnamefont
  {Coleman}}, \bibinfo {author} {\bibfnamefont {E.~A.}\ \bibnamefont {Mazenc}},
  \bibinfo {author} {\bibfnamefont {V.}~\bibnamefont {Shyam}}, \bibinfo
  {author} {\bibfnamefont {E.}~\bibnamefont {Silverstein}}, \bibinfo {author}
  {\bibfnamefont {R.~M.}\ \bibnamefont {Soni}}, \bibinfo {author}
  {\bibfnamefont {G.}~\bibnamefont {Torroba}},\ and\ \bibinfo {author}
  {\bibfnamefont {S.}~\bibnamefont {Yang}},\ }\Eprint
  {https://arxiv.org/abs/2110.14670} {arXiv:2110.14670 [hep-th]}  (\bibinfo
  {year} {2021})\BibitemShut {NoStop}%
\bibitem [{\citenamefont {Bzowski}\ and\ \citenamefont
  {Guica}(2019)}]{Bzowski:2018pcy}%
  \BibitemOpen
  \bibfield  {author} {\bibinfo {author} {\bibfnamefont {A.}~\bibnamefont
  {Bzowski}}\ and\ \bibinfo {author} {\bibfnamefont {M.}~\bibnamefont
  {Guica}},\ }\href {https://doi.org/10.1007/JHEP01(2019)198} {\bibfield
  {journal} {\bibinfo  {journal} {JHEP}\ }\textbf {\bibinfo {volume} {01}},\
  \bibinfo {pages} {198}},\ \Eprint {https://arxiv.org/abs/1803.09753}
  {arXiv:1803.09753 [hep-th]} \BibitemShut {NoStop}%
\bibitem [{\citenamefont {Ebert}\ \emph {et~al.}(2022)\citenamefont {Ebert},
  \citenamefont {Ferko}, \citenamefont {Sun},\ and\ \citenamefont
  {Sun}}]{toappear}%
  \BibitemOpen
  \bibfield  {author} {\bibinfo {author} {\bibfnamefont {S.}~\bibnamefont
  {Ebert}}, \bibinfo {author} {\bibfnamefont {C.}~\bibnamefont {Ferko}},
  \bibinfo {author} {\bibfnamefont {H.-Y.}\ \bibnamefont {Sun}},\ and\ \bibinfo
  {author} {\bibfnamefont {Z.}~\bibnamefont {Sun}},\ }\href@noop {} {\
  (\bibinfo {year} {2022})},\ \Eprint {https://arxiv.org/abs/22xx.xxxxx}
  {arXiv:22xx.xxxxx [hep-th]} \BibitemShut {NoStop}%
\bibitem [{\citenamefont {Taylor}(2018)}]{Taylor:2018xcy}%
  \BibitemOpen
  \bibfield  {author} {\bibinfo {author} {\bibfnamefont {M.}~\bibnamefont
  {Taylor}},\ }\Eprint {https://arxiv.org/abs/1805.10287} {arXiv:1805.10287
  [hep-th]}  (\bibinfo {year} {2018})\BibitemShut {NoStop}%
\bibitem [{\citenamefont {Hartman}\ \emph {et~al.}(2019)\citenamefont
  {Hartman}, \citenamefont {Kruthoff}, \citenamefont {Shaghoulian},\ and\
  \citenamefont {Tajdini}}]{Hartman:2018tkw}%
  \BibitemOpen
  \bibfield  {author} {\bibinfo {author} {\bibfnamefont {T.}~\bibnamefont
  {Hartman}}, \bibinfo {author} {\bibfnamefont {J.}~\bibnamefont {Kruthoff}},
  \bibinfo {author} {\bibfnamefont {E.}~\bibnamefont {Shaghoulian}},\ and\
  \bibinfo {author} {\bibfnamefont {A.}~\bibnamefont {Tajdini}},\ }\href
  {https://doi.org/10.1007/JHEP03(2019)004} {\bibfield  {journal} {\bibinfo
  {journal} {JHEP}\ }\textbf {\bibinfo {volume} {03}},\ \bibinfo {pages}
  {004}},\ \Eprint {https://arxiv.org/abs/1807.11401} {arXiv:1807.11401
  [hep-th]} \BibitemShut {NoStop}%
\bibitem [{\citenamefont {Ferko}\ \emph {et~al.}(2020)\citenamefont {Ferko},
  \citenamefont {Jiang}, \citenamefont {Sethi},\ and\ \citenamefont
  {Tartaglino-Mazzucchelli}}]{Ferko:2019oyv}%
  \BibitemOpen
  \bibfield  {author} {\bibinfo {author} {\bibfnamefont {C.}~\bibnamefont
  {Ferko}}, \bibinfo {author} {\bibfnamefont {H.}~\bibnamefont {Jiang}},
  \bibinfo {author} {\bibfnamefont {S.}~\bibnamefont {Sethi}},\ and\ \bibinfo
  {author} {\bibfnamefont {G.}~\bibnamefont {Tartaglino-Mazzucchelli}},\ }\href
  {https://doi.org/10.1007/JHEP02(2020)016} {\bibfield  {journal} {\bibinfo
  {journal} {JHEP}\ }\textbf {\bibinfo {volume} {02}},\ \bibinfo {pages}
  {016}},\ \Eprint {https://arxiv.org/abs/1910.01599} {arXiv:1910.01599
  [hep-th]} \BibitemShut {NoStop}%
\bibitem [{\citenamefont {Compere}\ \emph {et~al.}(2011)\citenamefont
  {Compere}, \citenamefont {McFadden}, \citenamefont {Skenderis},\ and\
  \citenamefont {Taylor}}]{Compere:2011dx}%
  \BibitemOpen
  \bibfield  {author} {\bibinfo {author} {\bibfnamefont {G.}~\bibnamefont
  {Compere}}, \bibinfo {author} {\bibfnamefont {P.}~\bibnamefont {McFadden}},
  \bibinfo {author} {\bibfnamefont {K.}~\bibnamefont {Skenderis}},\ and\
  \bibinfo {author} {\bibfnamefont {M.}~\bibnamefont {Taylor}},\ }\href
  {https://doi.org/10.1007/JHEP07(2011)050} {\bibfield  {journal} {\bibinfo
  {journal} {JHEP}\ }\textbf {\bibinfo {volume} {07}},\ \bibinfo {pages}
  {050}},\ \Eprint {https://arxiv.org/abs/1103.3022} {arXiv:1103.3022 [hep-th]}
  \BibitemShut {NoStop}%
\bibitem [{\citenamefont {Araujo-Regado}\ \emph {et~al.}(2022)\citenamefont
  {Araujo-Regado}, \citenamefont {Khan},\ and\ \citenamefont
  {Wall}}]{Araujo-Regado:2022gvw}%
  \BibitemOpen
  \bibfield  {author} {\bibinfo {author} {\bibfnamefont {G.}~\bibnamefont
  {Araujo-Regado}}, \bibinfo {author} {\bibfnamefont {R.}~\bibnamefont
  {Khan}},\ and\ \bibinfo {author} {\bibfnamefont {A.~C.}\ \bibnamefont
  {Wall}},\ }\href@noop {} {\  (\bibinfo {year} {2022})},\ \Eprint
  {https://arxiv.org/abs/2204.00591} {arXiv:2204.00591 [hep-th]} \BibitemShut
  {NoStop}%
\bibitem [{\citenamefont {Garc\'\i{}a}\ and\ \citenamefont
  {S\'anchez-Isidro}(2022)}]{Garcia:2022wad}%
  \BibitemOpen
  \bibfield  {author} {\bibinfo {author} {\bibfnamefont {J.~A.}\ \bibnamefont
  {Garc\'\i{}a}}\ and\ \bibinfo {author} {\bibfnamefont {R.~A.}\ \bibnamefont
  {S\'anchez-Isidro}},\ }\href@noop {} {\  (\bibinfo {year} {2022})},\ \Eprint
  {https://arxiv.org/abs/2209.06296} {arXiv:2209.06296 [hep-th]} \BibitemShut
  {NoStop}%
\end{thebibliography}%

\end{document}